\documentclass[
reprint,
superscriptaddress,
amsmath,
amssymb,
aps,
prb,
]{revtex4-2}
\usepackage{ragged2e}
\usepackage{caption}
\DeclareCaptionJustification{justified}{\justifying}
\usepackage[colorlinks,allcolors=cyan]{hyperref}
 
\usepackage{graphicx}
\usepackage{dcolumn}
\usepackage{bm}
\usepackage[capitalise]{cleveref}
\usepackage{physics}
\usepackage{xcolor}
\usepackage{comment}
\usepackage{bbold}
\usepackage{wrapfig}
\usepackage{mathtools}
\usepackage[caption=false]{subfig}
\DeclareCaptionOption{justified}[]{\caption@setjustification{justified}}

\newif\ifSM
\SMtrue
\newif\ifmainText
\mainTexttrue 

\begin{document}
\def\lc{\left\lfloor}   
\def\rc{\right\rfloor}
\setlength{\intextsep}{10pt plus 2pt minus 2pt}
\setlength{\abovedisplayskip}{4pt}
\setlength{\belowdisplayskip}{4pt}

\ifmainText
\title{Measurement induced criticality in quasiperiodic modulated random hybrid circuits}
 
\author{Gal Shkolnik}
 \affiliation{Racah Institute of Physics, The Hebrew University of Jerusalem, Jerusalem 91904, Israel}
 \author{Aidan Zabalo}
\affiliation{Department of Physics and Astronomy, Center for Materials Theory,
Rutgers University, Piscataway, New Jersey 08854, USA}
 \author{Romain Vasseur}
 \affiliation{Department of Physics, University of Massachusetts, Amherst, MA 01003, USA}
 \author{David A. Huse}
 \affiliation{Department of Physics, Princeton University, Princeton, New Jersey 08544, USA}
 
 \author{J. H. Pixley}
  \affiliation{Department of Physics and Astronomy, Center for Materials Theory,
Rutgers University, Piscataway, New Jersey 08854, USA}
\affiliation{Center for Computational Quantum Physics, Flatiron Institute, New York, New York 10010, USA}
 \author{Snir Gazit}
 \affiliation{Racah Institute of Physics, The Hebrew University of Jerusalem, Jerusalem 91904, Israel}
 \affiliation{The Fritz Haber Research Center for Molecular Dynamics, The Hebrew University of Jerusalem, Jerusalem 91904, Israel}
\date{\today}

\begin{abstract}
We study one-dimensional hybrid quantum circuits perturbed by quenched quasiperiodic (QP) modulations across the measurement-induced phase transition (MIPT). Considering non-Pisot QP structures, characterized by unbounded fluctuations, allows us to tune the wandering exponent $\beta$ to exceed the Luck bound $\nu \ge 1/(1-\beta)$ for the stability of the MIPT, where $\nu=1.28(2)$. Via robust numerical simulations of random Clifford circuits interleaved with local projective measurements, we find that sufficiently large QP structural fluctuations destabilize the MIPT and induce a flow to a broad family of critical dynamical phase transitions of the infinite QP type that is governed by the wandering exponent, $\beta$. We numerically determine the associated critical properties, including the correlation length exponent consistent with saturating the Luck bound, and a universal activated dynamical scaling with activation exponent $\psi \cong \beta$, finding excellent agreement with the conclusions of real space renormalization group calculations. 

\end{abstract}

\maketitle

{\it Introduction --} It is now well established that the competition between unitary evolution and local 
measurements in the evolution of generic quantum many-body systems
results in a dynamical phase transition in the properties of quantum trajectories, 
commonly termed the measurement-induced phase transition (MIPT) \cite{Nahum2019MIPT,Fisher2019MIPT,Fisher2018Zeno,Potter_2022,Fisher_2023}. 
The hallmark of this inherently out-of-equilibrium transition is a sharp transformation 
in the  structure of the wavefunction (or mixed density matrix)
between an entangling (mixed) and a disentangling (purifying) phase  in the respective limits of low and high measurement rates~\cite{Gullans2020Dynamical,Bao_2020,Jian_2020, Zabalo2020Critical,Nahum2019MIPT,Fisher2019MIPT}.

\begin{figure}
\begin{center}
    \includegraphics[width=0.35\textwidth, angle=270]{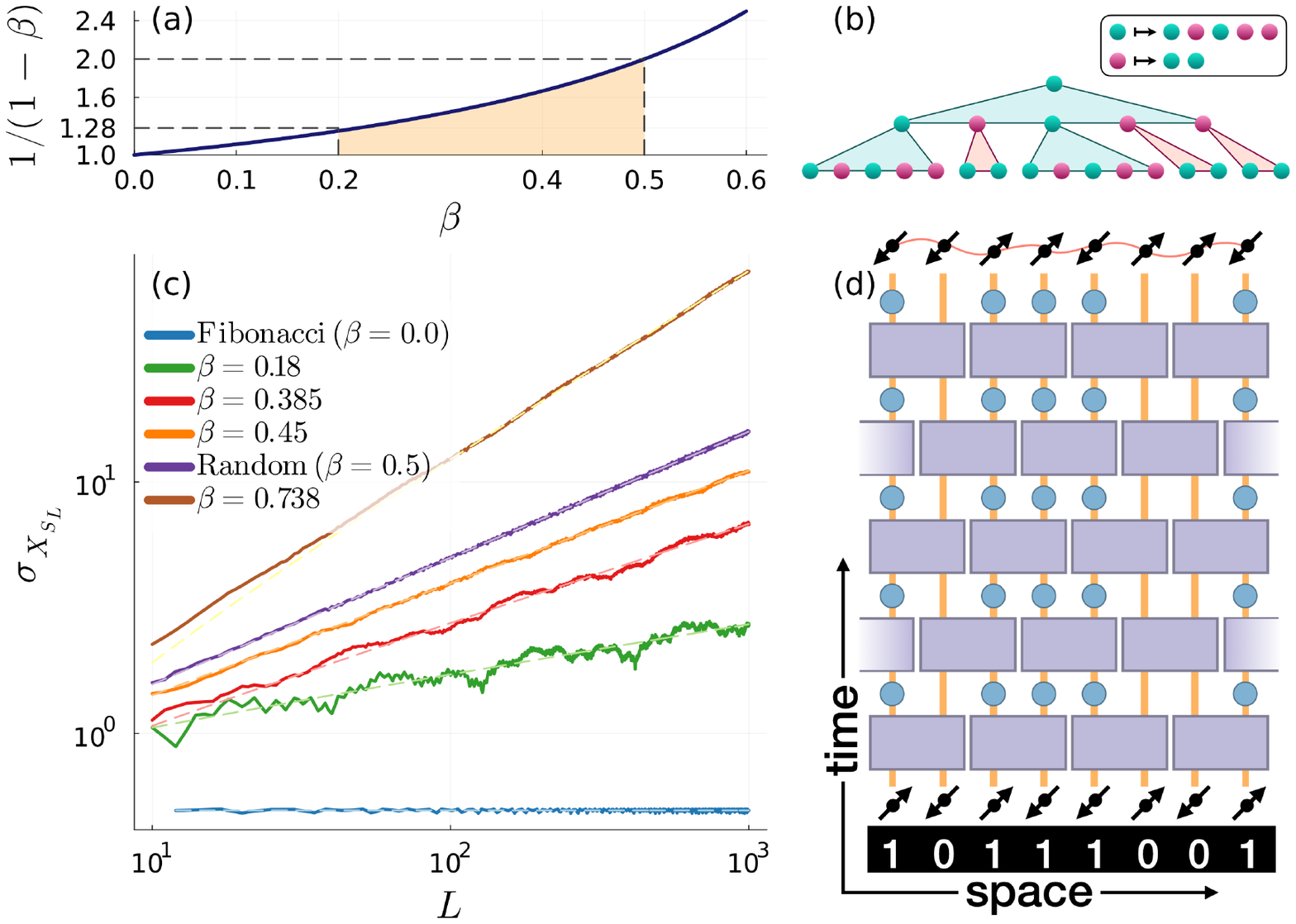}
  \end{center}
   \captionsetup[subfigure]{labelformat=empty}
    \subfloat[\label{subfig:LuckBound}]{}
    \subfloat[\label{subfig:SR}]{}
    \subfloat[\label{subfig:fluctutions}]{}
    \subfloat[\label{subfig:circuit}]{}
    \vspace*{-2mm}
  \caption{(a) Luck relevancy boundary. QP modulations are expected to destabilize the MIPT for $\beta\gtrsim 0.2$. Random disorder with $\beta=0.5$ has a minimal Luck stability bound of $\nu=2$. (b) Iterative construction of a QP chain by the substitution rule B in \cref{apdx:sub_rule}. $0$ ($1$) bits are marked by green (red) circles. (c) Scaling of structural fluctuations $\sigma_{X_{S_L}}$ as a function of the segment size $L$, shown in a log-log scale. Different curves correspond to several QP structures with varying wandering exponents $\beta$ and a random disorder ($\beta=1/2$). The displayed Fibonacci line contains only the local maxima. (d) Hybrid random circuit model. Purple rectangles represent the brick-wall structure of random two-qubit Clifford gates. Blue circles depict measurements, performed with probability $p$. The pattern of possible measurement is set by the QP structure, as described in the main text.}
  \label{fig:fig1_tmp}
  \vspace*{-2mm}
\end{figure}

Investigating physical properties of the MIPT has attracted a great deal of research interest, ranging from revealing what novel types of phases and phase transitions are possible \cite{lavasani2021measurement,PhysRevX.11.011030,PhysRevResearch.3.023200,BaoSymmetries,PhysRevX.12.041002,Zabalo2022InfiniteRand,Ippoliti2022fractal} to understanding the underlying universality class of the transition~\cite{Zabalo2020Critical,Bao_2020,Jian_2020, Zabalo-2022,Nahum2021Meas, Lu2021spacetime,nahum2023renormalization}, and of the entangling phase~\cite{PhysRevB.103.174309,PhysRevB.103.104306,PRXQuantum.4.010331}. 
These efforts have led to new insights, including mapping of the MIPT to classical statistical mechanics models in the replica limit \cite{Jian_2020,Bao_2020} with several interesting relations for free fermions and the ten-fold way~\cite{Cao2019fermion,Alberton2021Fermions,2020arXiv201204666J,Kells2023Topological,jian2023measurementinduced,fava2023nonlinear}, devising experimental probes \cite{Gullans_2020, Li2023cross}, and most notably, experimental studies of random hybrid circuit dynamics in ion-traps \cite{Noel2022exp} and in superconducting-based quantum processors \cite{koh2023measurement,hoke2023quantum}. 

Akin to static quantum phase transitions, it is important to understand the stability of the transition with respect to perturbations. Recent work \cite{Zabalo2022InfiniteRand} has shown that the relevance of quenched randomness can be understood through scaling theory, in particular the conventional picture of the Harris criterion $\nu \ge 2/d$, with $\nu$ being the correlation length exponent and $d$ the spatial dimension. As a result, for the conformally invariant MIPT~\cite{PhysRevB.104.104305, Gullans2020Dynamical} $\nu=1.28(2)$ for $d=1$, and, therefore, static disorder is a relevant perturbation to the MIPT, which drives a renormalization group  
``flow'' to a dynamical variant of the infinite-disorder phase transition, exhibiting ultra-slow activated dynamics (i.e., relaxation times scale as $\log(t^*)\sim \sqrt{L}$ for a system size $L$) and rare-events driven Griffiths phenomena. A natural question is whether other dynamical phase transitions exist between the conformally invariant MIPT and its flow to infinite randomness?

Immediate candidates are quasiperiodic modulations defined via deterministic yet aperiodic structures. In this setting, for weak perturbations, the stability of the critical point with correlation length exponent $\nu$ is determined by the Luck \cite{Luck1993classification} criterion 
\begin{equation}
\nu\geq1/{d\qty(1-\beta)}.
\end{equation} Here, $\beta$ is the wandering exponent, controlling the growth rate of the geometrical fluctuations with system size, as we explicitly define below. Conventional QP structures, such as the Fibonacci chain or an incommensurate cosine potential (with approximant wavevectors formed out of ratios of Fibonacci numbers), exhibit bounded fluctuations ($\beta=0$), which renders the perturbation irrelevant for the MIPT in one-dimension with $\nu=1.28(2)$ \cite{Nahum2019MIPT,Fisher2019MIPT,Gullans2020Dynamical}, as verified in \cite{Zabalo2022InfiniteRand} and in \cref{apdx:luck_irel}. By contrast, non-Pisot QP structures admit nontrivial wandering exponents, $\beta>0$, associated with {\it unbounded} fluctuations and hence can dramatically alter the above scenario. The fate of the MIPT in the presence of non-Pisot modulations is an outstanding question. 

In this paper, we study the influence of non-Pisot QP modulations on the MIPT via robust numerical simulations of one-dimensional random Clifford circuits competing with local projective measurements. Remarkably, we find that tuning the wandering exponent, $\beta$, renders QP a relevant perturbation in the RG sense and stabilizes a broad family of infinite QP fixed points, see \cref{subfig:LuckBound}. We numerically characterize the critical properties of this emergent transition and find that the correlation length exponent saturates the Luck bound. The dynamics follow a diverging dynamical exponent $z\to\infty$ with an activated scaling form governed solely by the wandering exponent, $\beta$. 
As a result, our work demonstrates how the MIPT universality class can be tuned to produce a range of entanglement scaling properties.

{\it Non-Pisot structures and universal behavior -- } We begin our discussion with a brief review of non-Pisot QP structures and associated critical behavior in static statistical mechanics models, see \cref{apdx:sub_rule} for further details. For concreteness, we generate binary QP structures by iteratively applying substitution rules that replace each $0$ ($1$) with a predefined binary string. 

More explicitly, initializing our bit string with a single bit $0$, at each step, we substitute every bit $0\backslash1$ with the bit strings $\varrho_{0\backslash1}$, e.g., see \cref{subfig:SR}. The substitution rules encoded in $\varrho_{0\backslash1}$ uniquely determine the resulting QP structure, and for our needs, a crucial property is the wandering exponent, $\beta$ \cite{Luck1993classification}, defined as follows: For each contiguous segment of digits $S_L$ of length $L$, we write $X_{S_L}=\sum_{i\in S_L} x_i$, with $x_i$ being the digit at position $i$. The standard deviation of $X_{S_L}$ scales as $\sigma_{X_{S_L}} \sim L^{d\beta}$ (up to subleading corrections), where averaging is carried out over all segments $S_L$, belonging to infinitely large QP bit strings.

For randomly drawn bits, the standard random walk diffusive scaling gives $\beta=1/2$. The Fibonacci chain has a vanishing wandering exponent $\beta=0$, hence admits bounded fluctuations. By contrast, non-Pisot structures define a family of QP chains with non-trivial wandering exponents $\beta>0$ and unbounded fluctuations \cite{Luck1993classification,Godreche1992, Hermisson_1997,hermisson2000aperiodic}, as depicted in Fig.~\ref{subfig:fluctutions}. As mentioned, this allows pushing the wandering exponent beyond the stability of the conformally invariant MIPT, $1/(1-\beta)> \nu \approx 1.28$.

To understand the resulting critical properties for the MIPT, we follow a similar line of reasoning as outlined for the random case~\cite{Zabalo2022InfiniteRand}. By employing the replica trick in the limit of an infinite onsite Hilbert space dimension, Refs.~\cite{Bao_2020,Jian_2020,Li2021StatMapping} have established a mapping between hybrid random circuit dynamics and the statistical mechanics of a static two-dimensional, $S_{Q!}$-symmetric classical Potts model, with $S_{Q!}$ being the permutation group of $Q!$ elements, and $Q$ the number of replicas. In this picture, the MIPT transition corresponds to the standard order-disorder transition of the Potts model in the replica limit $Q\to 1$. In particular, volume and area-law phases are identified with the ferromagnetic and paramagnetic phases of the Potts model, respectively. 

Introducing quenched QP modulations with unbounded fluctuations to the measurement probabilities translates to coupling constants modulations of the dual quantum Potts model. Importantly, we can still carry out a real space renormalization group (RSRG) analysis for any replica index that can be analytically continued to the physical replica limit~\cite{Zabalo2022InfiniteRand}. This establishes an equivalence with the low energy properties of the quantum Potts chain with QP-modulated couplings. 

Applying the QP variant of the RSRG treatment to our problem shows that at the critical point of an infinite QP type~\cite{Luck1993classification,Hermisson_1997,hida2004new,hermisson2000aperiodic,InfiniteQPNatComm}, the long-time dynamical relaxation rate $t^*$ has an activated space-time scaling form
\begin{equation}
\label{eqn:NPansatz}
   \log(t^*) \sim L^\psi,
\end{equation}
for the chain length $L$ and a universal activation exponent $\psi$. For infinite QP fixed points, it has been found that $\psi=\beta$ \cite{Hermisson_1997,hermisson2000aperiodic} whereas for the infinite randomness fixed point $\psi=1/2$ \cite{Fisher1995Random,Young_1996}.

The above mapping has direct implications for the scaling of the steady-state long-time bi-partite entanglement entropy. We note that within the statistical mechanics picture, the entanglement entropy is proportional to the free energy cost of a boundary domain wall, which translates into the expected scaling relation $S\sim L^\beta$, in analogy with the random case with $\beta=1/2$~\cite{Zabalo2022InfiniteRand}. In the following, we test the above predictions, derived in the replica limit, with numerically exact simulations of the physically relevant cases of qubits under hybrid random Clifford circuit dynamics.

{\it QP modulated quantum circuits and observables --} We consider one-dimensional hybrid quantum circuits of linear size $L$. The dynamics alternates between unitary evolution generated by random two-qubit Clifford gates and local projective measurements in the computational ($z$) basis, see \cref{subfig:circuit}. Crucially, the measurement rate, $p(x)$ is static but {\it position-dependant} following a non-Pisot QP structure in real space. The measurement probability at site $i$ is then defined as $p(i)=p$ if $x_i=1$ and $p(i)=0$ otherwise. Throughout the paper, we consider periodic boundary conditions. Since the QP segments are not periodic (in the system size), the boundary hosts a defect and could be an additional source of finite size effects. 

Similarly to the uniform case, the measurement rate $p$ serves as a tuning parameter that controls the relative strength of the measurement process compared to the unitary dynamics. Expectation values are evaluated by the combined ensemble averaging over random circuits and QP structure realizations. To that end, we initially generate a long non-Pisot QP bit string at the order of $10^8$ digits, from which we randomly cut contiguous segments of length $L$ digits. The presented value of $p$ is normalized by the measurement gates density, see \cref{apdx:sub_rule}.

To investigate physical properties differentiating the distinct phases and probe the putative critical properties, we define the following observables. As is standard, the bi-partite von-Neumann entanglement entropy $S(A)$ for a pure wavefunction $\ket{\psi}$ and a segment $A$ is obtained by tracing over the complementary set of sites $A^C$. This allows defining the reduced density matrix $\rho_A\coloneqq\Tr_{A^C}|\psi\rangle\langle\psi|$ from which we evaluate the associated entanglement entropy $S(A)\coloneqq-\Tr_A\left[\rho_A\log_2\rho_A\right]$.

To eliminate subleading finite size scaling effects, it is beneficial to inspect the behavior of the tri-partite mutual information \cite{Zabalo2020Critical}, defined as
\begin{equation}
\begin{split}
    \mathcal{I}_3(A,B,&C) \coloneqq S(A)+S(B)+S(C)-S(A\cup B)\\ & -S(A\cup C)-S(B\cup C)+S(A\cup B\cup C),
\end{split}
\end{equation}
where we consider a partition of the chain into adjacent segments $A, B, C$ of size $L/4$. This combination precisely cancels out the boundary contributions in the area-law phase and turns negative with a linear scaling $\mathcal{I}_3\sim L$ in the volume-law phase. 
To extract critical properties, we consider $P[\mathcal{I}_3=0]$ instead of $\expval{\mathcal{I}_3}$ due to the broad distribution of the latter. Following the above argument, $P[\mathcal{I}_3=0]$ is expected to approach unity in the area-law phase and vanish in the volume-law phase.

\begin{figure}
\begin{center}
    \includegraphics[width=0.45
\textwidth]{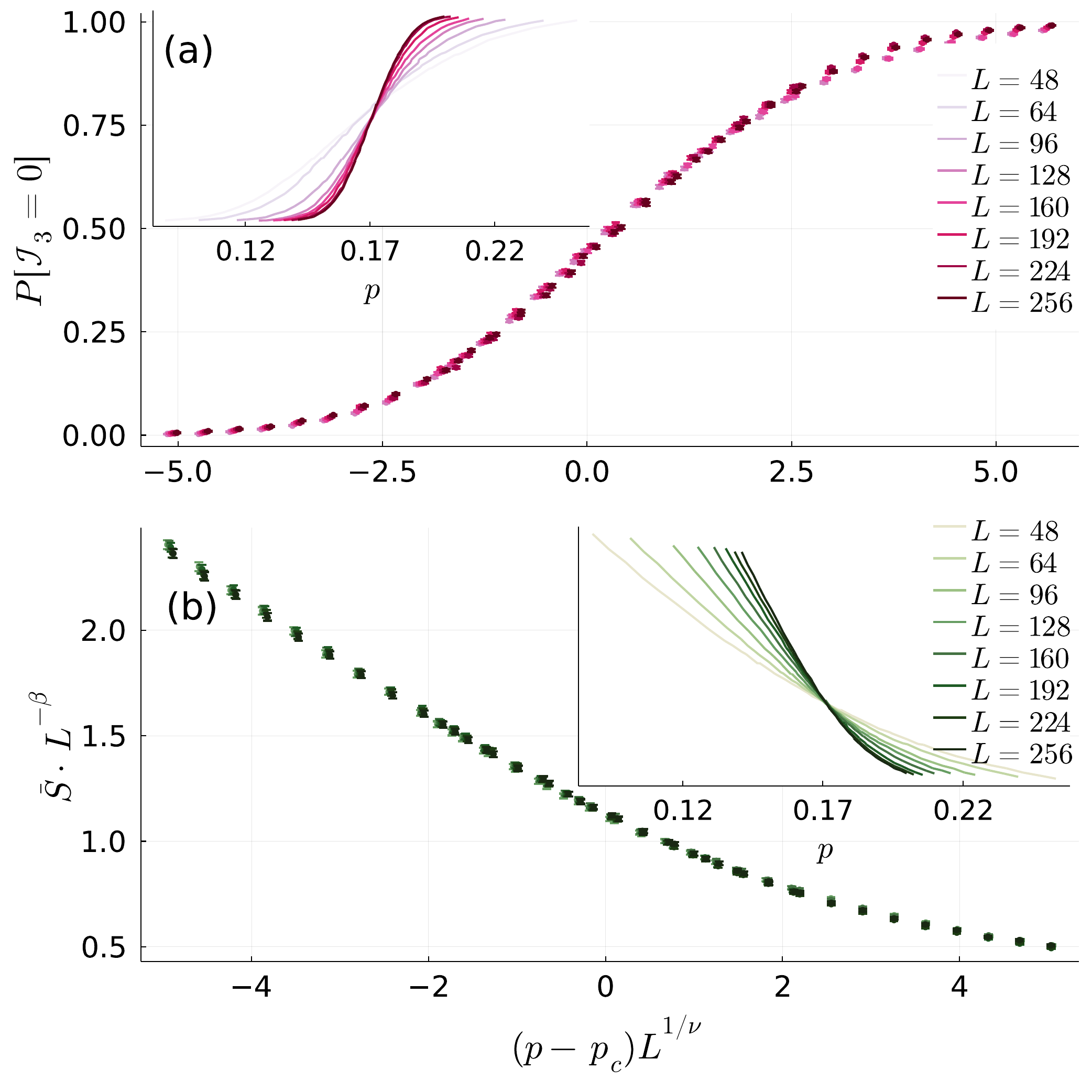}
  \end{center}
   \captionsetup[subfigure]{labelformat=empty}
    \subfloat[\label{subfig:PI3}]{}
    \subfloat[\label{subfig:Sbeta}]{}
    \vspace*{-2mm}
  \caption{Curve collapse analysis of (a) $P\left[\mathcal{I}_3=0\right]$ and (b) $S/L^{\beta}$, using the scaling variable $y=(p-p_c)L^{1/\nu}$, for $p_c=0.169(2)$ and $\nu=1.62$. Insets depict the data prior to scaling. Different curves corresponding to different system sizes collapse. The data shown here and in \cref{fig:fig3_tmp} are for the non-Pisot structure B \cref{apdx:sub_rule}.}
  \label{fig:fig2_tmp}
  \vspace*{-2mm}
\end{figure}

Lastly, we investigate the dynamics of an ancilla qubit, initialized in a Bell state with a qubit located at the middle of the chain. We carry out a random unitary evolution (without measurements) for a duration $T=2L$, which entangles the ancilla qubit with the rest of the system. Following that, we employ the full hybrid dynamics defined above. We then assess the purification time at which the ancilla qubit disentangles from the chain, as can be probed via the entanglement entropy $S_{Q}$. 

{\it Numerical results and scaling properties --} We focus on a non-Pisot QP structure with $\beta=0.385$, defined by the substitution rule, $0\mapsto 01011$ and $1\mapsto 0$, labeled by B in \cref{apdx:sub_rule}. Our first task is to determine the critical measurement rate $p_c$ in the presence of non-Pisot QP modulations. With that goal in mind, we track the evolution of $P\qty[\mathcal{I}_3=0]$ as a function of the measurement rate, shown in the inset of \cref{subfig:PI3}. We observe that $P[\mathcal{I}_3=0]$ vanishes deep in the volume-law phase, for $p\to0$, and in the complementary limit, $p\to1$, it approaches unity as expected in the area-law phase. 

Furthermore, we find that curves of the universal amplitude $P[\mathcal{I}_3=0]$ corresponding to an increasing set of linear system sizes cross at a single point, indicating a continuous transition. This allows us to carry out a curve collapse analysis using the ansatz $P[\mathcal{I}_3=0]\sim g\qty((p - p_c)L^{1/\nu})$. In our fitting procedure, we utilize the Luck stability criterion to impose a lower bound on $\nu\geq 1/d\left(1-\beta\right)\approx 1.626$. We obtained consistent results without the bound. This gives our numerical estimate for the critical measurement rate $p_c=0.169(2)$ and $\nu=1.6(1)$, shown in \cref{subfig:PI3}, leading to an excellent curve collapse. Interestingly, our numerical findings suggest that $\nu$ precisely saturates the Luck bound. 

\begin{figure}
\begin{center}
    \includegraphics[width=0.45\textwidth]{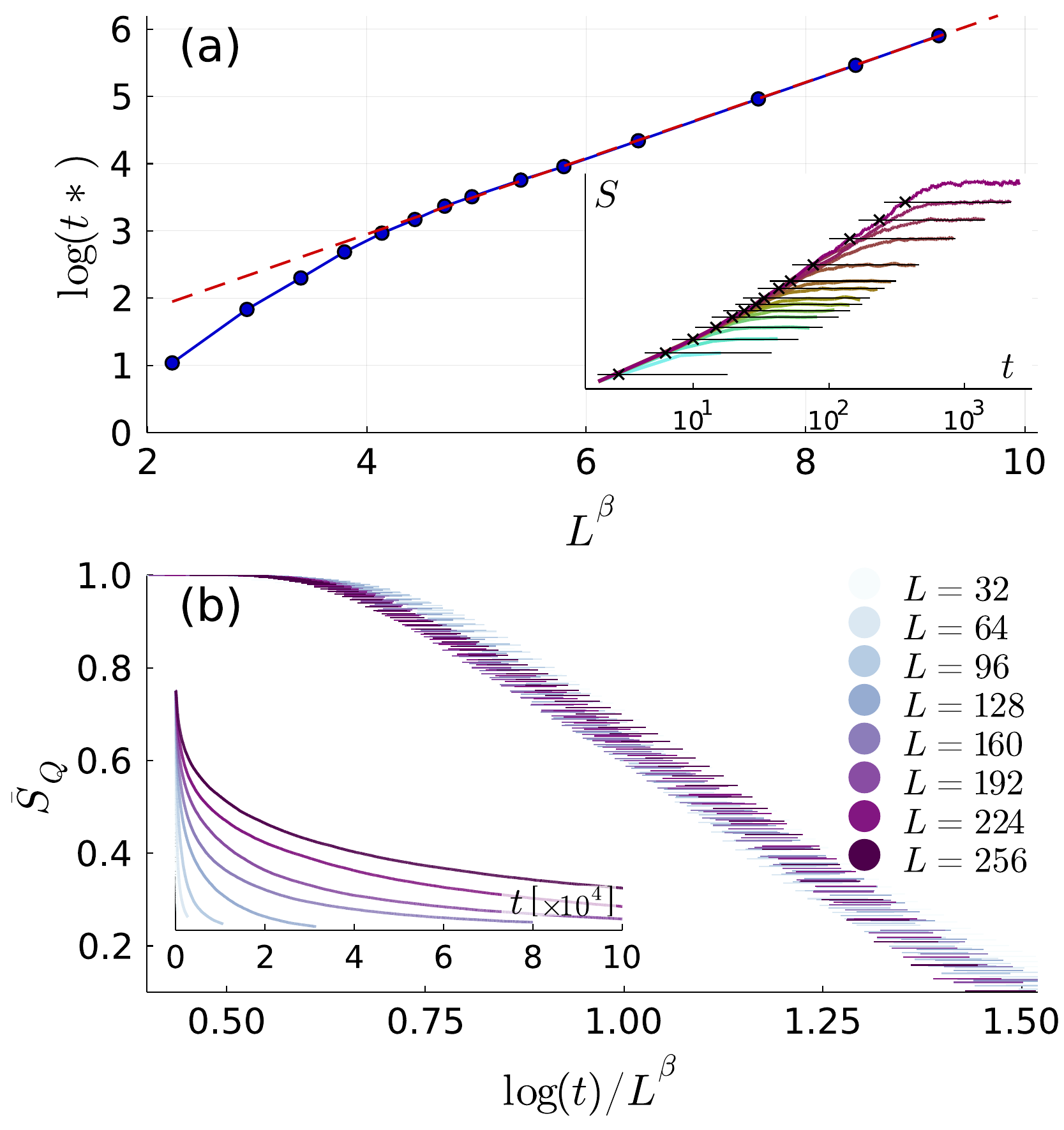}
  \end{center}
  \captionsetup[subfigure]{labelformat=empty}
    \subfloat[\label{subfig:tsat_Lbeta}]{}
    \subfloat[\label{subfig:ancilladyn}]{}
    \vspace*{-2mm}
  \caption{(a) Logarithm of the saturation time $t^*$ of the half-cut entanglement entropy, $S(L/2)$, at criticality versus $L^\beta$. The dashed red curve is a linear fit. The inset shows the saturation time extraction method. Crosses represent $t^*$ for each system size $L$. (b) Activated dynamics scaling of the ancilla order parameter, $S_Q(t,L)$, at criticality. Different curves correspond to different system sizes. Curve collapse is obtained via the scaling variable $\log(t)/L^{\beta}$. See \cref{apdx:sub_rule} for uncertainty calculation.
  \vspace*{-2mm}
  }
  \label{fig:fig3_tmp}
\end{figure}
        
To further corroborate the above results, we investigate the critical properties of the saturated half-cut bi-partite entanglement entropy, $S\qty(L/2)$, obtained at long times. Following the above scaling arguments, we consider the scaling hypothesis
\begin{equation}
    S\qty(L/2,t\rightarrow\infty)\sim L^\beta f\qty(\qty(p-p_c)L^{1/\nu}),
\end{equation}
and hence the ratio $S/L^{\beta}$ is predicted to have a vanishing scaling dimension, as depicted at the inset of \cref{subfig:Sbeta}. Indeed, we identify a crossing of curves corresponding to different system sizes at the same critical measurement rate $p_c=0.1700(3)$. This key result provides numerical support for the entanglement entropy scaling predictions. Moreover, the scaling variable $y=(p - p_c)L^{1/\nu}$ yields an excellent curve collapse, with $\nu = 1.65(2)$, in agreement with our previous estimate using a different observable. 

Next, we turn to study critical dynamical properties. To test the predicted activated scaling behavior, \cref{eqn:NPansatz}, at criticality ($p=p_c$), we estimate the saturation time, $t^*$, at which $S\qty(L/2,t)$ reaches its late time value \cite{Zabalo2022InfiniteRand}, see the inset of \cref{subfig:tsat_Lbeta}. To highlight the activated scaling form, in \cref{subfig:tsat_Lbeta}, we depict $\log(t^*)$ as a function of $L^\beta$. The resulting linear dependence at large $L$ values precisely agrees with the dynamical scaling ansatz. 

Motivated by the above result, we further inspect the universal dynamics through the evolution of the ancilla order parameter, $S_Q(t,L)$, at $p_c$ \cite{Gullans_2020,Zabalo2020Critical}, shown in the inset of \cref{subfig:ancilladyn}. Conveniently, $S_Q$ admits a vanishing scaling dimension and hence should follow the scaling form
\begin{equation}
\label{eqn:ancillascaling}
    {S}_{Q}(t,L)\sim f\qty(\log(t)/L^{\beta}).
\end{equation}
To test the above functional dependence, in \cref{subfig:ancilladyn}, we plot the temporal axis using the scaled variable $\log(t)/L^\beta$, we indeed observe the expected curve collapse for sufficiently large $L$. 

Away from the critical point, activated scaling in random systems is typically associated with power law space-time scaling, characterized by a varying dynamical exponent due to Griffiths singularities~\cite{Griffiths_1969,Fisher1995Random}. The underlying mechanism for this effect is rare events, which are not expected in deterministic structures such as QP modulations. Indeed, within the available system sizes, our numerical results likely indicate the absence of a similar effect for the studied QP modulations, as we describe in \cref{apdx:rare_reg}, though we do see the appearance of a cross-over at finite time that cuts off the relaxation time $\log\qty(t^*(p)) \sim \xi^\psi\sim |p-p_c|^{-\nu \psi}$.

We performed the above analysis for several additional values of $\beta$, see \cref{apdx:luc_rel,apdx:luck_irel}. We found that $\beta$ values satisfying the Luck criterion maintain the same critical properties of the pristine MIPT. On the other hand, when the Luck criterion is violated, we identified the same universal behavior presented above, which is characterized by an activated dynamical scaling and saturation of the Luck bound, only with a modified $\beta$. 

{\it Discussion and summary -- } We uncovered an infinite-QP criticality driven by non-Pisot QP modulations of the MIPT. Remarkably, relevant wandering exponents, as defined by the Luck criterion, stabilize a broad family of critical points whose scaling properties are controlled solely by the corresponding $\beta$ values, namely we obtain $\nu = 1/(1-\beta)$, spatial entanglement grows like $S \sim L^{\beta},$ and the activation exponent $\psi=\beta$. This observation affords a rather non-trivial generalization of the random case precisely given by $\beta=1/2$, studied in \cite{Zabalo2020Critical}. Another interesting aspect is the equivalent influence of QP modulations in the low energy properties of the static quantum Potts and the long-time evolution of the hybrid circuit dynamics. This linkage further supports the utility of the replica trick representation of MIPT in predicting out-of-equilibrium critical properties \cite{Bao_2020,Jian_2020,Li2021StatMapping}.

Looking to the future, the wandering exponent can also be tailored in random systems by introducing correlations \cite{Crowley2019hyperuniform}. Contrasting its effect with the QP case presented in our work would be interesting. Another research direction is exploring a generalized phase diagram resulting from the competition between QP modulations and quenched random disorder.  In particular, based on the above scaling arguments, for $\beta<1/2$, quenched disorder is expected to overwhelm QP and induce a flow to an infinite-disorder fixed point, while for $\beta>1/2$, QP is expected to remain stable. We leave these outstanding questions to future studies. 

\acknowledgements{
We thank Sarang Gopalakrishnan, Michael Gullans, and Justin Wilson for collaborations on related work. This work is supported in part by the BSF Grant No. 2020264 (G.S., J.H.P., S.G.),  the Alfred P. Sloan Foundation through a Sloan Research Fellowship (R.V., J.H.P.), the Air Force Office of Scientific Research under Grant No. FA9550-21-1-0123 (R.V.), and the Office of Naval Research grant No.  N00014-23-1-2357 (J.H.P.).  D.A.H. was supported in part by (U.S.A.) NSF QLCI grant OMA-2120757.  G.S. acknowledges the hospitality of the Center for Materials Theory at Rutgers University visitor program and the support of the Council for Higher Education Scholarships Program for Outstanding Doctoral Students in Quantum Science and Technology. This work was performed in part at the Aspen Center for Physics, which is supported by National Science Foundation grant PHY-2210452 (J.H.P.).
}

\appendix
\section{Substitution rule generation of QP structures}
\label{apdx:sub_rule}

For simplicity, we consider binary substitution rules that map each digit to a finite length bit string, explicitly defined by
\begin{equation}
    \begin{cases}
        &0\mapsto \varrho_0\\
        &1\mapsto \varrho_1
    \end{cases}.
\end{equation}
The QP structure is constructed via a recursive application of the above substitution rule. In practice, we generate QP arrays of order $10^8$ bits. Specific realizations are contiguous segments whose initial index is drawn uniformly. The substitution rules studied in this work are summarized in \cref{tab:subrule}.

\begin{table}[b]
\begin{tabular}{| c | c | m{1.8cm} | m{1.9cm} |} \hline
Label & Substitution rule & Wandering exponent & Measurement density \\ [0.5ex] 
 \hline\hline
 Fibonacci & $\begin{cases}
                        &0\mapsto 01\\
                        &1\mapsto 0
                    \end{cases}$ & 0.0 & 0.382 \\ 
 \hline
 A & $\begin{cases}
            &0\mapsto 11\\
            &1\mapsto 1100
        \end{cases}$ & 0.180 & 0.618 \\ 
 \hline
 B & $\begin{cases}
            &0\mapsto 01011\\
            &1\mapsto 00
        \end{cases}$ & 0.385 & 0.451 \\
 \hline
 C & $\begin{cases}
            &0\mapsto 011\\
            &1\mapsto 01000
        \end{cases}$ & 0.450 & 0.414 \\
 \hline
\end{tabular}
\vspace*{15mm}
\centering
\caption{\label{tab:subrule}Substitution rules and corresponding properties of the QP chains considered in this work.}
\end{table}

Some geometrical properties of the QP structure can then be computed from the characteristic matrix of $\varrho$,
\begin{equation}
    M_{\varrho}\coloneqq
    \begin{pmatrix}
        \#_0\varrho_0 & \#_0\varrho_1 \\
        \#_1\varrho_0& \#_1\varrho_1
    \end{pmatrix},
\end{equation}
where $\#_{0}\varrho_1$ counts the number of $0$'s in the substitution rule of $1$, and similarly for all other cases. The largest eigenvalue of $M_{\varrho}$, $\lambda_1$, determines the asymptotic inflation factor per substitution. Its corresponding eigenvector is proportional to the digit densities. In the main text, we use this information to normalize the QP modulated measurement rate, such that it directly compares to the uniform hybrid circuit.

The second-largest eigenvalue, $\lambda_2$, controls the digit density fluctuations. In particular, when $|\lambda_2|<1$ fluctuations are bounded, while for $|\lambda_2|>1$ they diverge as a power-law with a wandering exponent $\beta$ \cite{Godreche1992},
\begin{equation}
\label{eqn:log_periodic}
    \sigma\qty[X_{S_L}]\sim G\qty(\frac{\log L}{\log \lambda_1}) L^{\beta},\;\beta=\frac{\ln{|\lambda_2|}}{\ln{\lambda_1}}.
\end{equation}
Here $G$ is a log-periodic function.

\begin{figure}
\begin{center}
    \includegraphics[width=0.48\textwidth]{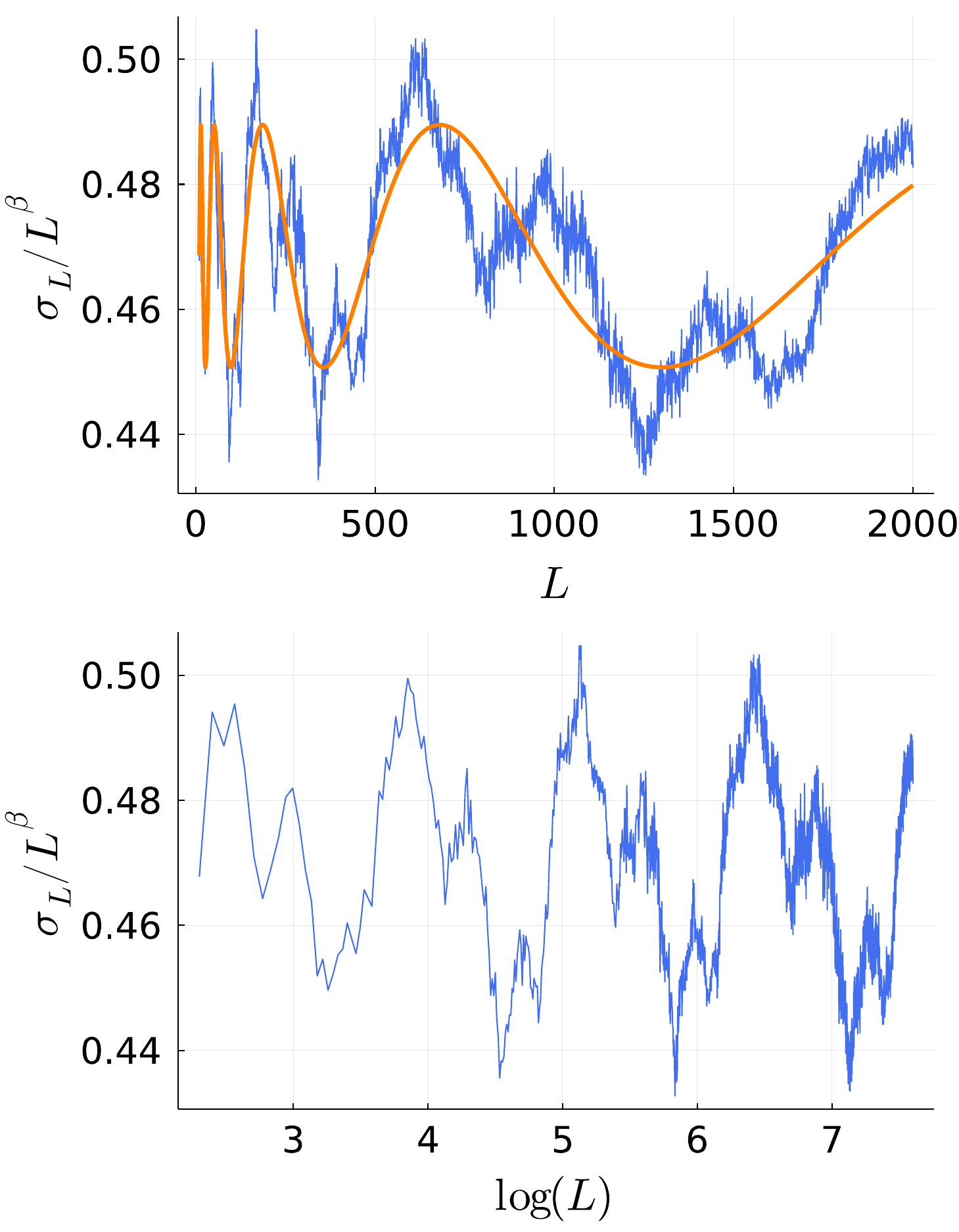}
  \end{center}
  \vspace*{7mm}
  \caption{The log-periodic function $G$,  \cref{eqn:log_periodic}, associated with the substitution rule $B$, \cref{tab:subrule}, with $\beta=0.385$. The log-periodicity (orange curve) agrees well with the theoretical prediction $\log \lambda_1\approx \log\qty(1+\sqrt{7})$.}
  \label{fig:logperiod}
\end{figure}

For the most part, the log-periodic function $G\qty(\log(L)/\log(\lambda_1))$ has little effect on scaling properties such that the simpler form $\sigma\qty[X_{S_L}]\sim L^{\beta}$ provide an excellent approximation. However, we observed that the critical dynamics of the ancilla order parameter exhibit log-periodic modulations. To estimate this effect, we consider a generalized scaling function in \cref{eqn:ancillascaling} in the main text that takes into account the full form of geometrical fluctuations, including the log-periodic modulations, $S_{Q}\sim f(\log(t)/\qty[G\qty(\log(L)/\log(\lambda_1))L^{\beta}])$. $G$ is nowhere differentiable \cite{Godreche1992}, and hence can not be written in a closed form. Instead, we model $G$ as a source of uncertainty in the scaling variable (horizontal axis) of \cref{subfig:ancilladyn}. in the main text. By examining its functional dependence of $L$, see \cref{fig:logperiod}, we can provide a rough bound $0.44<G<0.5$, which we translate to an uncertainty region, $\delta x\approx 0.06 x$, where $x=\log(t)/\qty(0.47L^{\beta})$. This reasoning is used in our horizontal error estimates in \cref{eqn:ancillascaling} in the main text of the main text.

\section{$\mathcal{I}_3$ analysis}
\label{apdx:I3}

In this section, we provide a detailed analysis of the scaling properties of the tri-partite mutual information, $\mathcal{I}_3$. First, we examine the full distribution $P\qty[\mathcal{I}_3]$. In \cref{subfig:I3dist_L256}, we fix $L=256$, and depict $P\qty[\mathcal{I}_3]$ for several values of $p$. In the area-law phase, $p>p_c$, the distribution is localized about $0$. For the complementary limit, $p<p_c$, in the volume-law phase, the distribution progressively broadens and attains a finite mean value. These observations are exemplified in \cref{subfig:I3dist_vol} and \cref{subfig:I3dist_area}. The above trend is evident as we increase the system size. 

\begin{figure}[h!]
\begin{center}
    \includegraphics[width=0.5\textwidth]{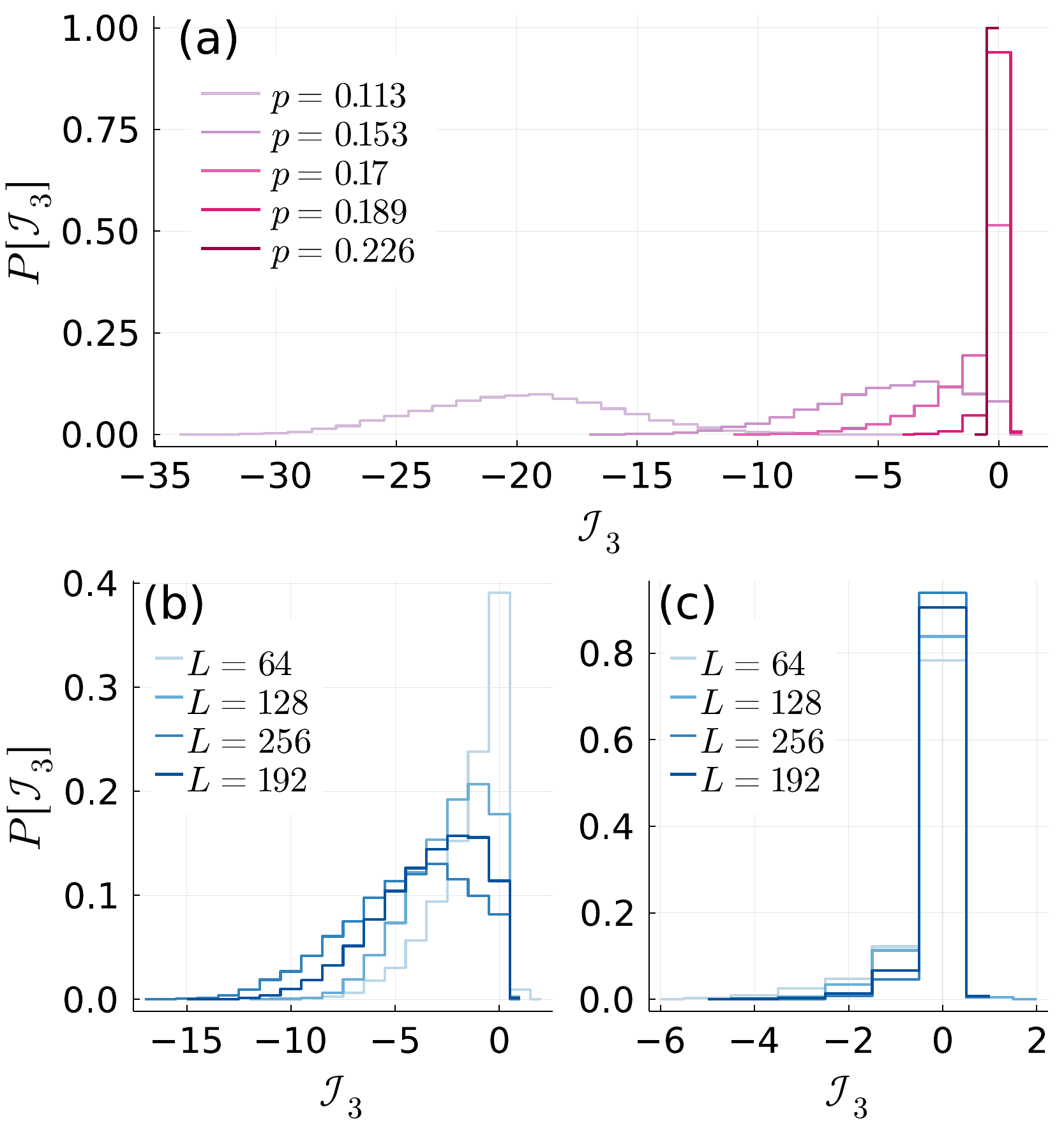}
  \end{center}
  \captionsetup[subfigure]{labelformat=empty}
    \subfloat[\label{subfig:I3dist_L256}]{}
    \subfloat[\label{subfig:I3dist_vol}]{}
    \subfloat[\label{subfig:I3dist_area}]{}
  \caption{Distribution of $\mathcal{I}_3$ for (a) a system of size $L=256$ and different values of $p$. (b) Different system sizes in the volume-law phase, $p\approx0.153$. (c) Different system sizes in the area-law phase, $p\approx0.189$.
  }
  \vspace*{3mm}
  \label{fig:I3dist}
\end{figure}

Next, we consider the critical properties of $\expval{\mathcal{I}_3}$, which can, in principle, admit a nontrivial dimension, leading to the scaling ansatz
\begin{equation}
    \mathcal{I}_3\sim L^{\eta}f\qty((p-p_c)L^{1/\nu}).
\end{equation}

\begin{figure}[h!]
  \begin{center}
    \includegraphics[width=0.5\textwidth]{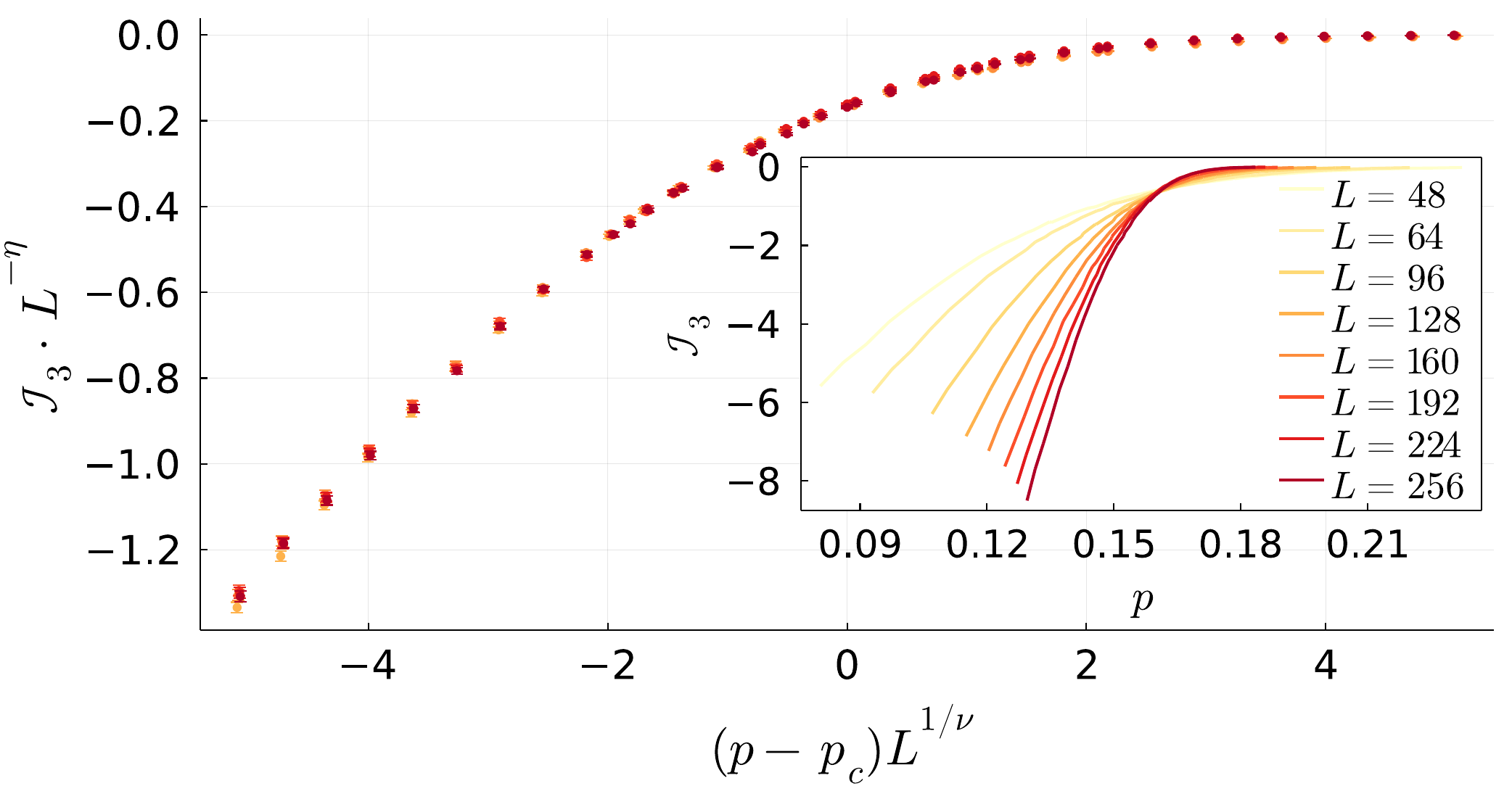}
  \end{center}
  \vspace*{7mm}
  \caption{Curve collapse analysis of $\expval{\mathcal{I}_3}$, with $p_c$ fixed according to previous analysis, yielding $\nu=1.64(5)$ and $\eta=0.34(2)$. The inset shows data prior to scaling.}
  \label{fig:I3pt_pc}
\end{figure}

In our curve collapse analysis, we fix $p_c=0.17$, as computed using a different observable, and consider $\nu,\eta$ as unknowns in our fitting procedure. This gives an excellent curve collapse with our numerical estimate of $\eta=0.34(2)$ and $\nu=1.64(5)$, see \cref{fig:I3pt_pc}.

\section{Luck-Relevant QP modulations}

\label{apdx:luc_rel}
\begin{figure}[hbt!]
\begin{center}
    \includegraphics[width=0.5\textwidth]{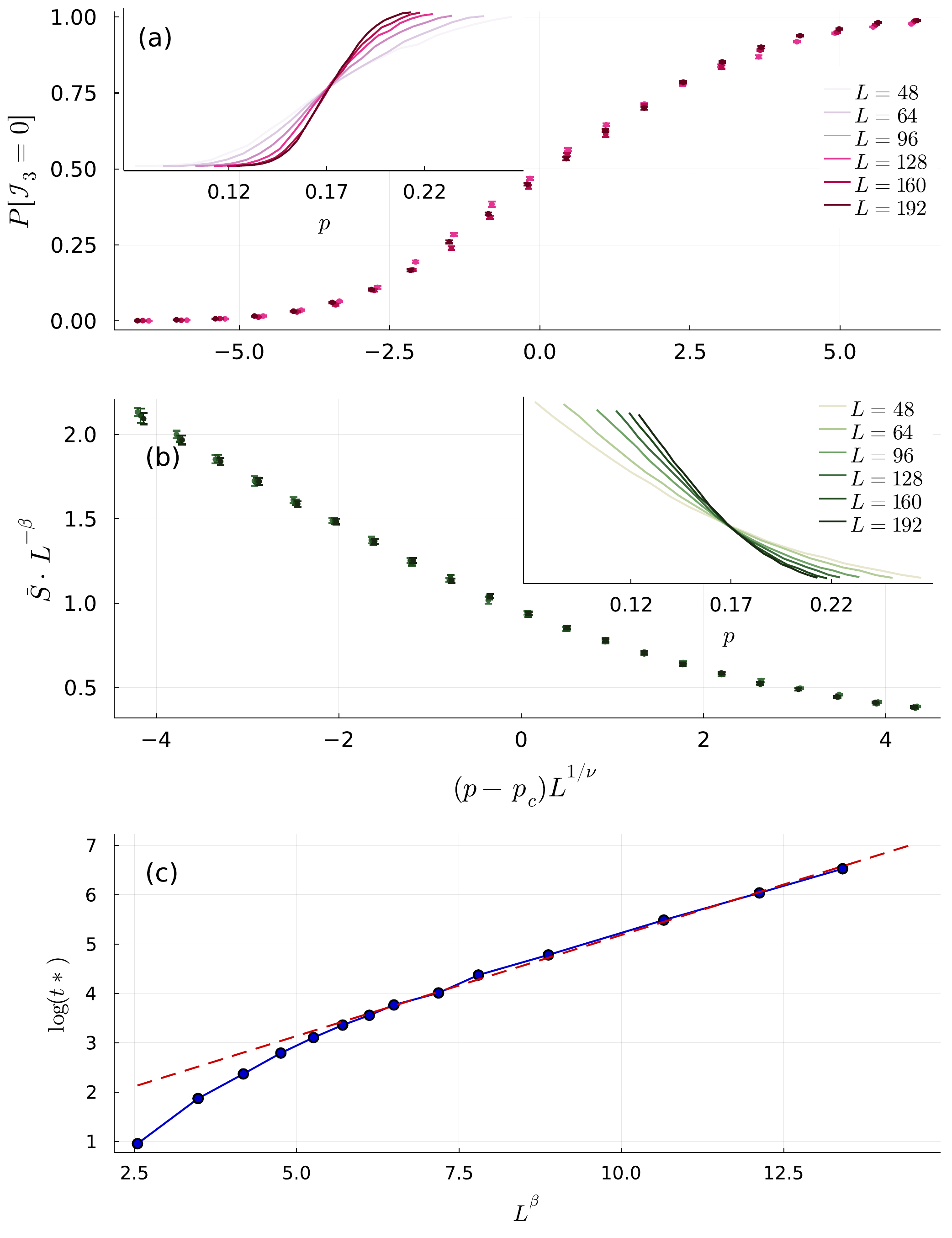}
  \end{center}
  \vspace*{-3mm}
  \captionsetup[subfigure]{labelformat=empty}
    \subfloat[]{}
    \subfloat[]{}
    \subfloat[\label{subfig:dyn0.450}]{}
  \caption{Critical finite size scaling analysis of the substitution rule C, \cref{tab:subrule}, with $\beta=0.450$. Curve collapse analysis of (a) $P[\mathcal{I}_3=0]$ and (b) $S/L^{\beta}$ using the scaling variable $(p-p_c)L^{1/\nu}$. In (c), we demonstrate the activated dynamical scaling of the saturation time $t^*$. Similarly to the substitution rule B, we find a linear dependence (dashed red line) between $\log(t^*)$ and $L^{\beta}$.}
  \vspace*{2mm}
  \label{fig:0.450}
\end{figure}

In this section, we repeat the analysis carried out in the main text for a different substitution rule (C in \cref{tab:subrule}) characterized by the wandering exponent $\beta\approx0.450$. The corresponding Luck bound is $\nu\geq 1.82$, and hence it presents a relevant perturbation for the MIPT. Specifically, in \cref{fig:0.450}, we present a curve collapse analysis of $P[\mathcal{I}_3=0]$. This yields $p_c=0.169(8)$ and $\nu=1.8(5)$, which, importantly, saturates the lower bound set by Luck criterion. Furthermore, the  predicted universal amplitude $S/L^{\beta}$ display a curve collapse using  $p_c=0.174(1)$ and $\nu=1.84(7)$, in agreement with the previous estimate. Lastly, we investigate critical dynamics by tracking the evolution of the saturation time at $p_c$ as a function of system size, see \cref{subfig:dyn0.450}. Indeed, we observe the predicted activated scaling with $\psi=\beta$ for sufficiently large system sizes. The above results are fully consistent with the ones obtained for substitution rule B, in the main text, only with a different wandering exponent.

\section{Luck-irrelevant QP modulations}
\label{apdx:luck_irel}

\begin{figure}[h]
\begin{center}
     \includegraphics[width=0.5\textwidth]{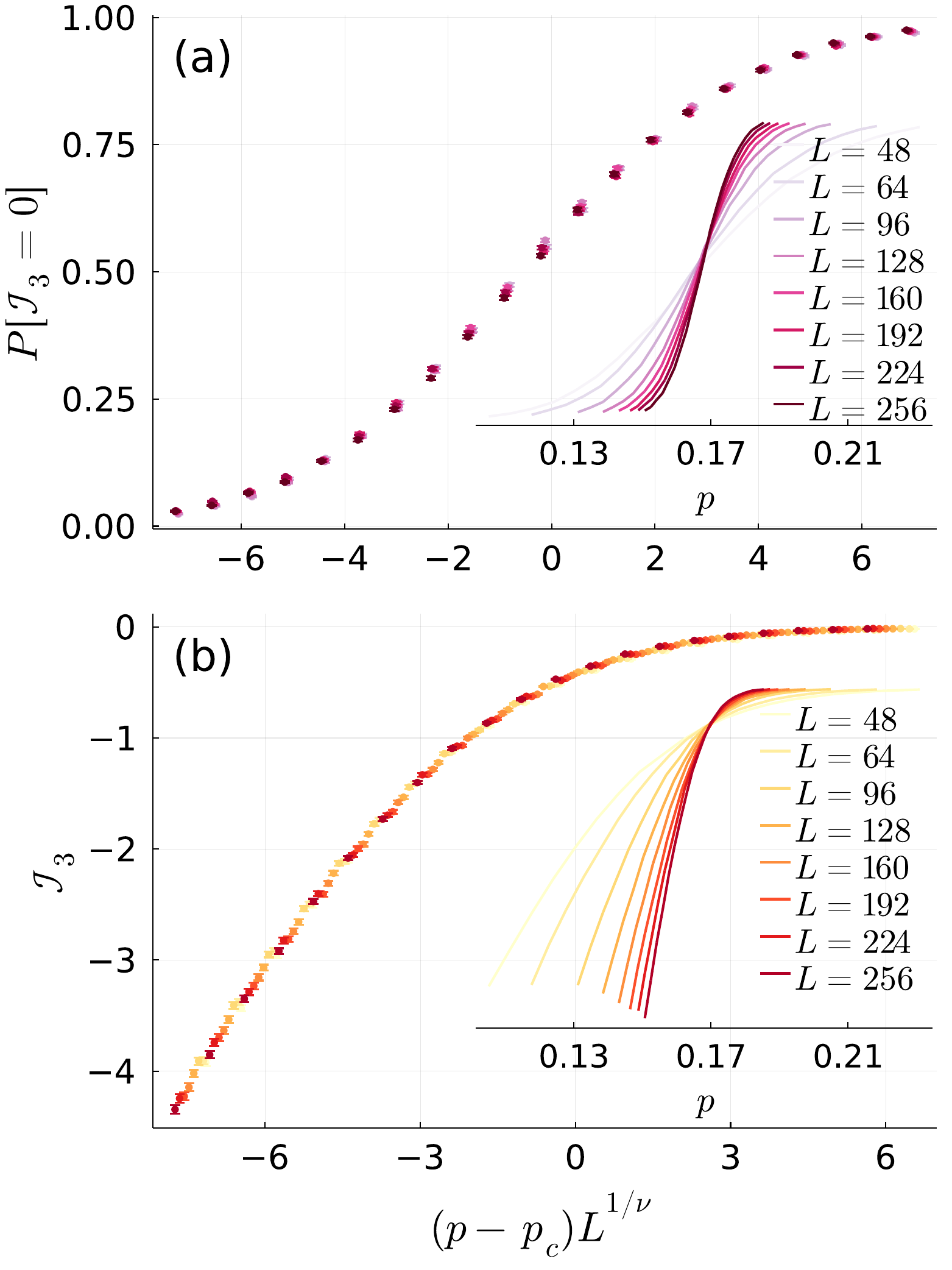}
  \end{center}
  \vspace*{5mm}
  \captionsetup[subfigure]{labelformat=empty}
  \caption{Finite size scaling analysis of substitution rule A, with the irrelevant $\beta=0.180$. Curve collapse analysis of (a) $P[\mathcal{I}_3=0]$ and (b) $\expval{\mathcal{I}_3}$. Insets show the data prior to scaling.
  }
  \vspace*{5mm}
  \label{fig:pt0.180}
\end{figure}

The Luck bound suggests that QP modulations are irrelevant for sufficiently small wandering exponents, $\beta<0.2$. However, this bound is derived for weak modulations. Binary QP patterns, following sharp modulation, are generally not a weak perturbation. Therefore, their effect should be studied numerically. To that end, below, we analyze two additional QP structures. The first follows the substitution rule A in \cref{tab:subrule}, with a finite wandering exponent $\beta=0.180$ that is Harris-Luck irrelevant. The second is the substitution rule generated by the Fibonacci chain, which has bounded fluctuations, \cref{tab:subrule}.
\begin{figure}[hbt!]
\begin{center}
    \includegraphics[width=0.5\textwidth]{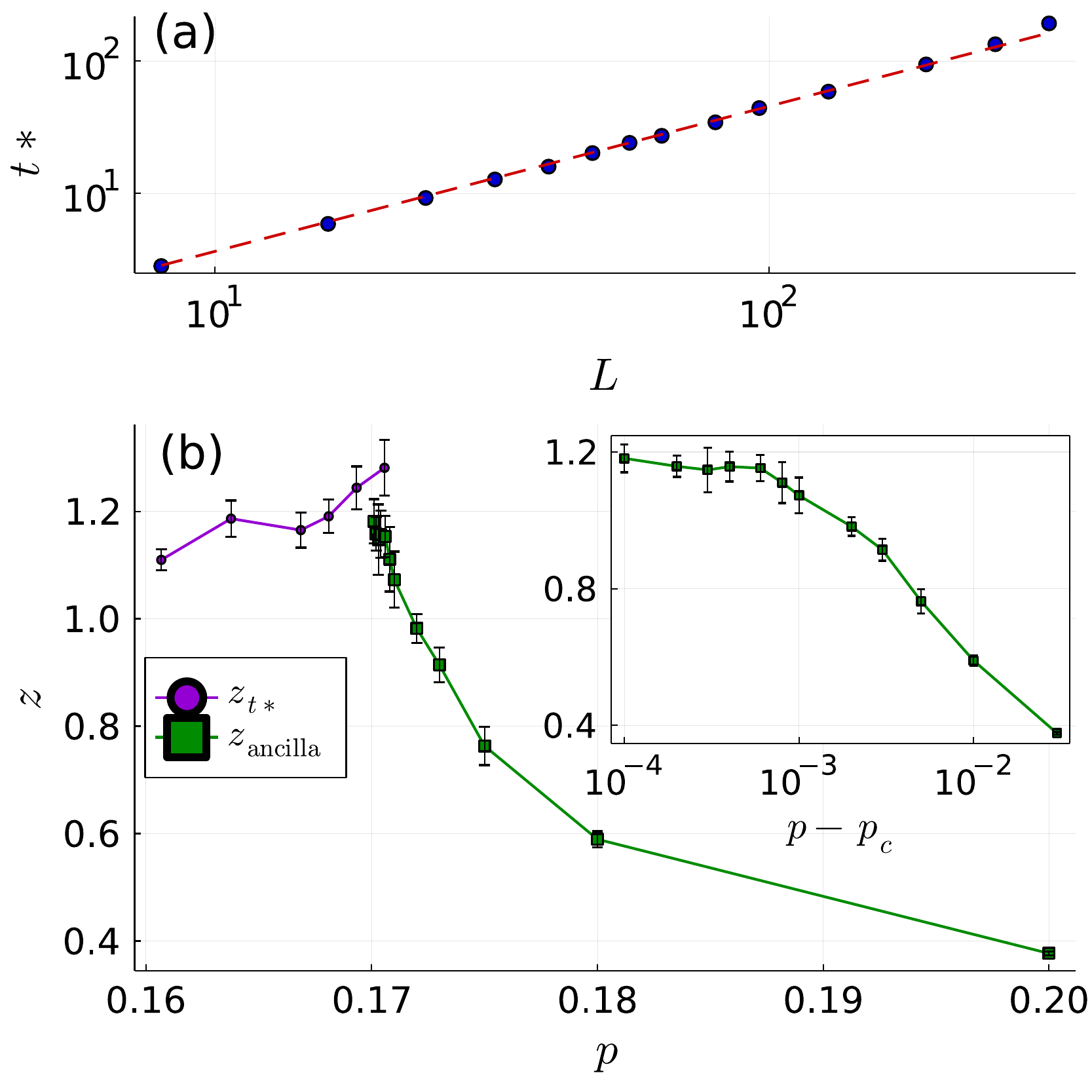}
  \end{center}
  \captionsetup[subfigure]{labelformat=empty}
  \subfloat[\label{subfig:dyn_0.180_crit}]{}
  \subfloat[\label{subfig:dyn_0.180_scan}]{}
  \caption{(a) The saturation time, $t^*$, at criticality versus the system size. A fit to a power law form $t^*\sim L^z$ gives $z=1.091(8)$ in agreement with the relativistic scaling of the MIPT. (b) The dynamical exponent $z$, as determined by the saturation time, $t^*$, and the ancilla purification time, as a function of the measurement rate $p$ in the vicinity of the critical point $p_c$. The inset focuses on the critical regime.
  }
  \label{fig:dyn0.180}
  \vspace*{5mm}
\end{figure}

We begin our analysis with $\beta=0.180$. Our curve collapse results of $P[\mathcal{I}_3=0]$ and $\expval{\mathcal{I}_3}$ yield $p_c=0.1686(2),\,\nu=1.31(3)$ and $p_c=0.1708(2),\,\nu=1.324(9)$ with a vanishing scaling dimension $\eta=0$, respectively, see \cref{fig:pt0.180}. The resulting $\nu$ value conforms with the unperturbed MIPT, suggesting the irrelevance of this structure.

In addition, the critical dynamics appears to follow the relativistic scaling of the unperturbed MIPT. In particular, we find an approximate linear growth of saturation time with the lattice size, with a dynamical exponent $z\approx 1.091(8)$, see \cref{subfig:dyn_0.180_crit}. The small deviation from the relativistic result, $z=1$, is likely due to finite-size effects. Moreover, in \cref{subfig:dyn_0.180_scan,subfig:dyn_0.180_scan}, we estimate $z$ in the vicinity of the critical regime using both the saturation time (in the volume-law phase) and the ancilla purification time (in the area-law phase). We, indeed, find that $z$ remains bounded and approximately equals unity.

\begin{figure}[hbt!]
\begin{center}
    \includegraphics[width=0.5\textwidth]{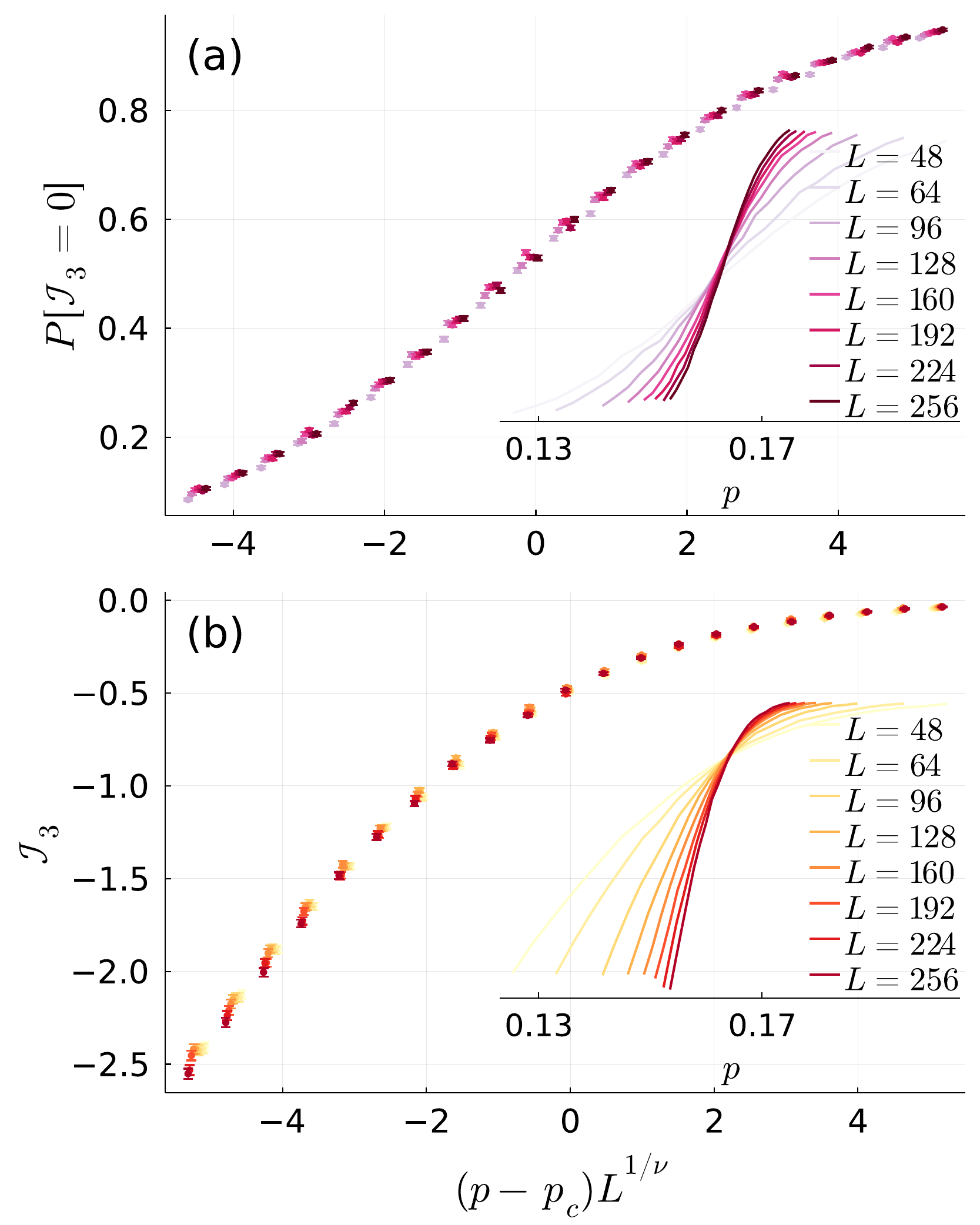}
  \end{center}
  \captionsetup[subfigure]{labelformat=empty}
    \subfloat[\label{subfig:fib_pI3}]{}
    \subfloat[\label{subfig:fib_avI3}]{}

  \caption{Finite size scaling analysis of the Fibonacci chain. Curve collapse analysis of (a) $P[\mathcal{I}_3=0]$, (b) $\expval{\mathcal{I}_3}$. Insets show the data prior to scaling.}
  \label{fig:fibonacci}
  \vspace*{5mm}
\end{figure}

Finally, we study the effect of binary Fibonacci chain modulations. The smooth (cosine) modulations were shown to be irrelevant in \cite{Zabalo2022InfiniteRand}. The corresponding curve collapse analysis provides the estimates $p_c=1.6314(9), \nu=1.29(2)$ for $P[\mathcal{I}_3=0]$, see \cref{subfig:fib_pI3}. our curve collapse analysis of $\mathcal{I}_3$ yields $p_c=0.16440(8), \nu=1.264(7)$, and $\eta=0$ see \cref{subfig:fib_avI3}. The dynamical exponents fully agree with relativistic scaling, as obtained by fitting the saturation growth with systems size that gives $z=1.050(5)$, in \cref{fig:fib_tsat}. From above results, we can conclude that binary Fibonacci chain modulations are irrelevant at the MIPT. 

\begin{figure}[hbt!]
\begin{center}
    \includegraphics[width=0.5\textwidth]{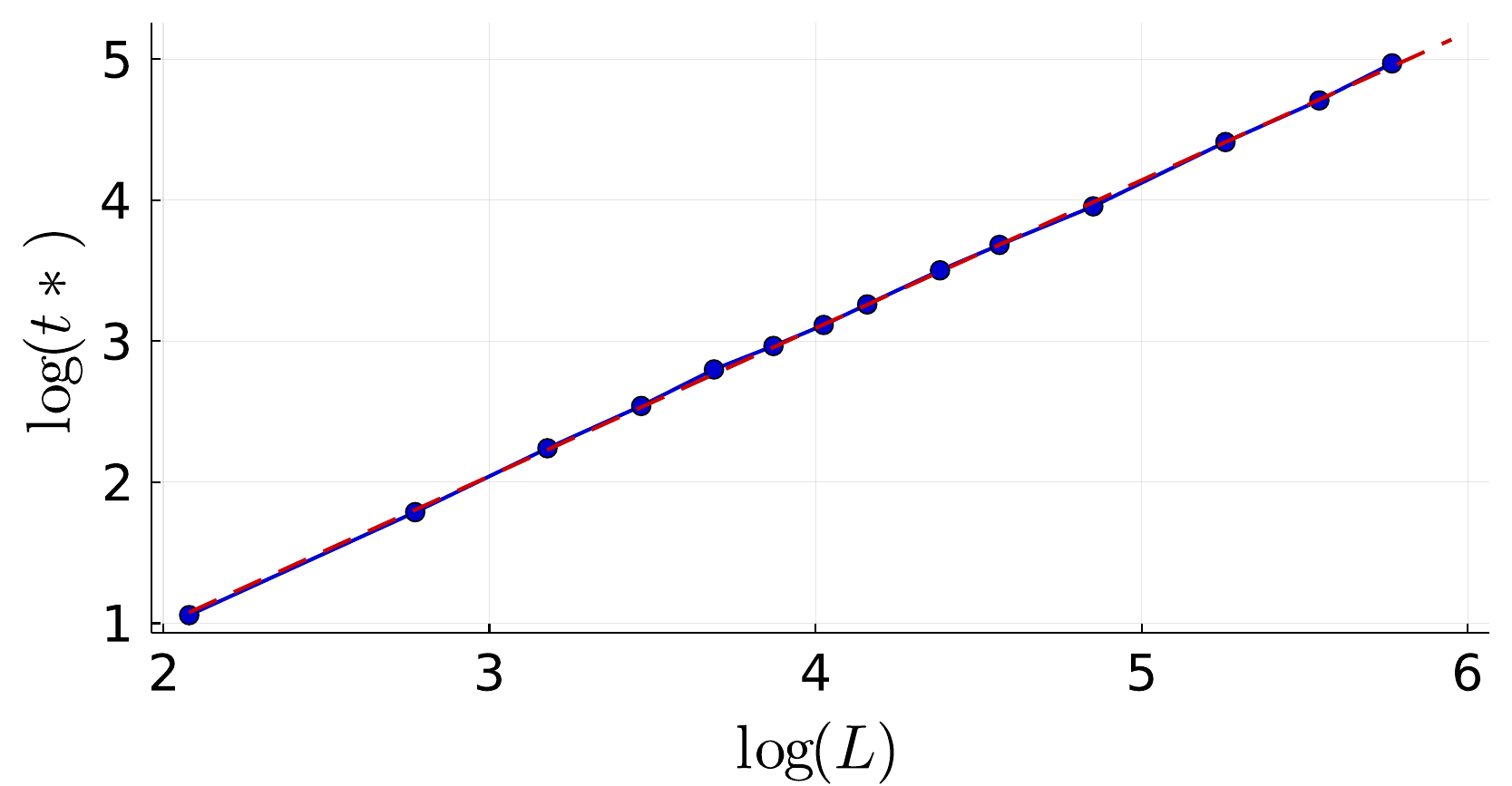}
  \end{center}
\vspace*{5mm}
  \caption{The saturation time $t^*$ displays a linear (dashed red line) growth with L.}
  \label{fig:fib_tsat}
\end{figure}

\section{Rare regions and Griffiths effect} \label{apdx:rare_reg}
\begin{figure}[b!]
\begin{center}
    \includegraphics[width=0.5\textwidth]{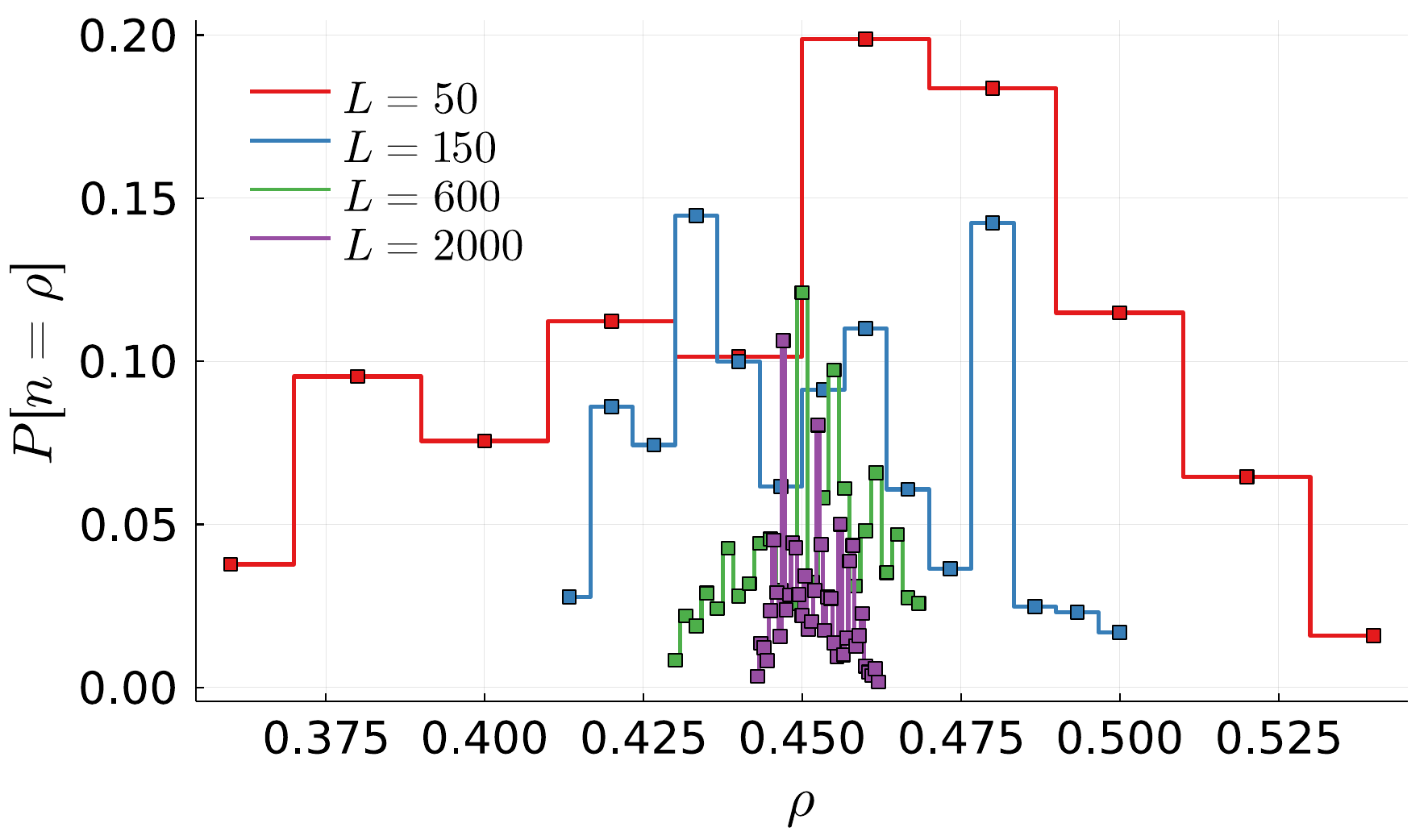}
  \end{center}
  \captionsetup[subfigure]{labelformat=empty}
  \subfloat[\label{}]{}
  \caption{The distribution of $n\coloneqq X_{S_L}/L$, for various segments of length $L$, for the substitution rule $B$.
  }
  \label{fig:ones_density}
\end{figure}

As mentioned in the main text, for infinite disorder transitions, activated dynamical scaling is typically accompanied by Griffiths singularities \cite{Vojta_2019,Griffiths_1969,Zabalo2022InfiniteRand} with a varying dynamical exponent that progressively diverges upon approach to criticality. The underlying mechanism is local atypical disorder configurations that belong to the opposite global phase of matter. While the probability of realizing such rare events is exponentially suppressed with the chain length (for some arbitrary $c$), $w \sim e^{-cL}$, it competes with the exponentially small energy splitting between distinct ordered ground states in order-disorder transitions. This competition results in power-law singularities away from criticality, in particular, a critical divergence of the dynamical exponent  $1/z\sim |p-p_c|L^{\psi\nu}$ \cite{Fisher1995Random,Hooyberghs2004}, with $\nu$ being the correlation length exponent and $\psi$ the activation exponent.

\begin{figure}[h!]
\begin{center}
    \includegraphics[width=0.5\textwidth]{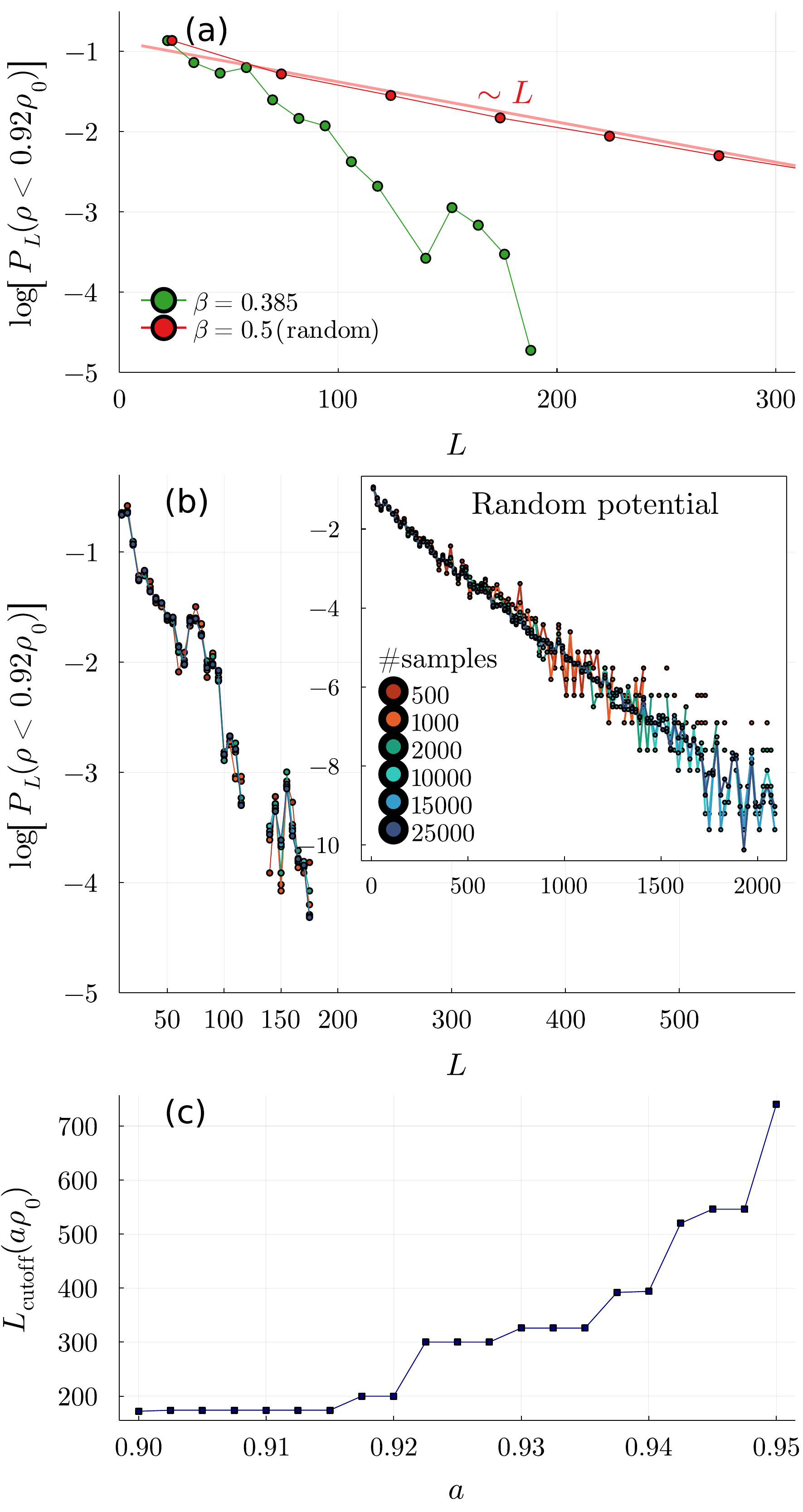}
  \end{center}
  \captionsetup[subfigure]{labelformat=empty}
    \subfloat[\label{subfig:rareeventsdecay_a}]{}
    \subfloat[\label{subfig:rareeventsdecay_b}]{}
    \subfloat[\label{subfig:rareeventsdecay_c}]{}
  \caption{(a) Log of the probability of finding a segment of length $L$ with ones density lower than a cutoff of $a\rho_0$, with $a=0.92$, $P_L[\rho<0.92\rho_0]$, for the substitution rule 
 B (green line) and random disorder (red line). (b) Behavior of the distribution tails, $P_L[\rho<0.92\rho_0]$. The main panel (inset) displays data for substitution rule B (random disorder). Different curves correspond to different numbers of samples. (c) Lattice size above which $P_L[\rho<0.92\rho_0]$ identically vanishes as a function of the cutoff $a$.
  }
  \label{fig:rareeventsdecay}
\end{figure}

Returning to our case, in the following, we investigate whether the non-Pisot QP structures, studied in this work, support Griffiths singularities. To that end, we examine the probability density of rare events in these modulations.
In \cref{fig:ones_density}, we display the measurement gates density distribution, $P[n=\rho]$, with $n\coloneqq X_{S_L}/L$ for an increasing set of system sizes $L$. The distribution is centered on the mean density, $\rho_0=0.451$, for structure B, \cref{tab:subrule}. Importantly, for each system size, we observe a sharp drop, above which the probability density identically vanishes. This indicates the absence of long tails associated with rare events.

\begin{figure}[h!]
\begin{center}
    \includegraphics[width=0.5\textwidth]{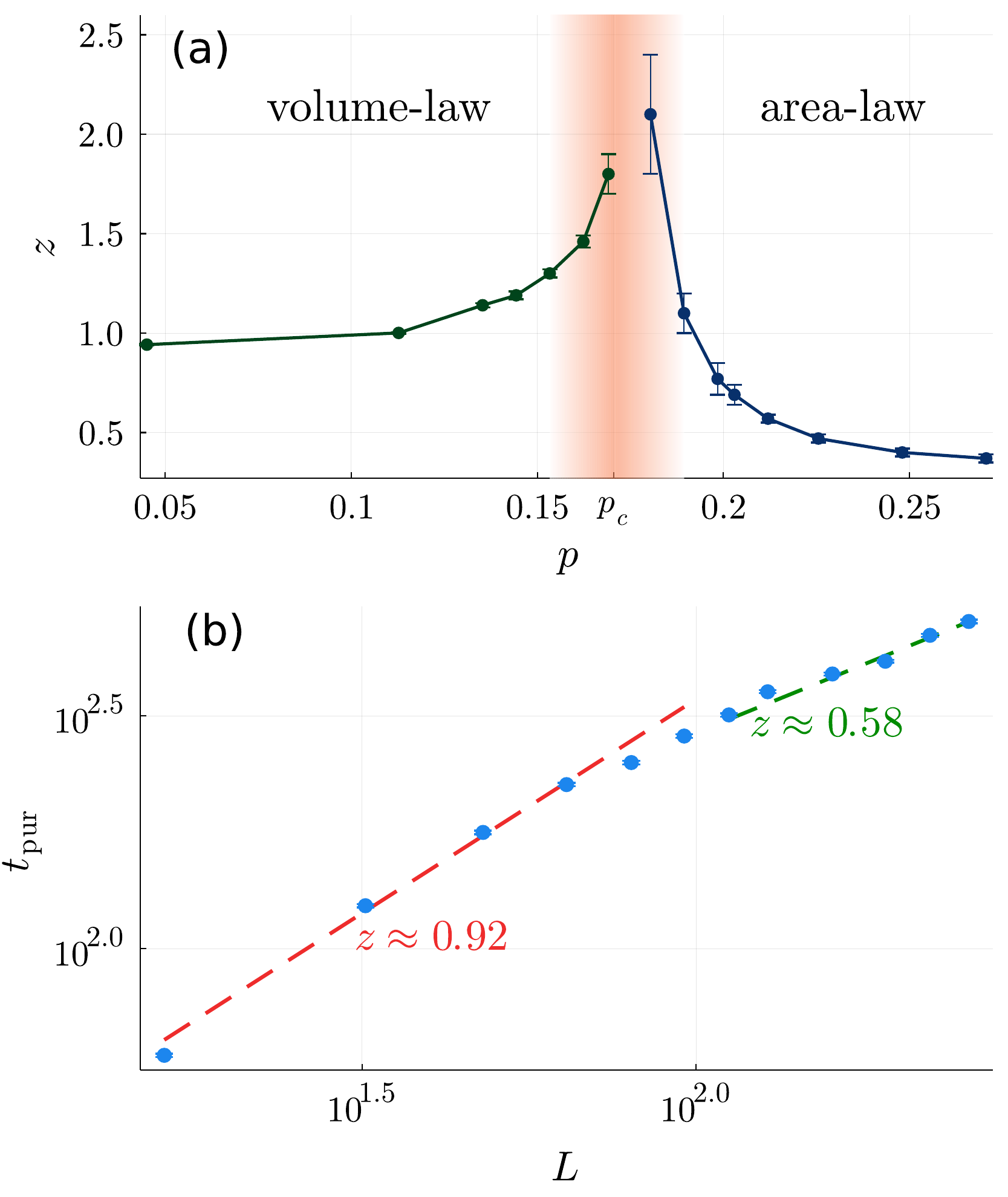}
  \end{center}
  \captionsetup[subfigure]{labelformat=empty}
    \subfloat[\label{subfig:cross_over_a}]{}
    \subfloat[\label{subfig:cross_over_b}]{}
  \caption{(a) The effective dynamical exponents $z$, defined by a power-law fit for the substitution rule B as a function of the measurement rate across the MIPT. The typical time in the volume-law phase is the entanglement entropy saturation time, and in the area-law phase is the ancilla qubit purification time. The fit considers only system sizes $L\geq 64$ . (b) The ancilla qubit purification time as a function of the chain length $L$ for the measurement rate  $p=0.212$ on a log-log scale. The red (green) dashed line is a power-law fit to the data points $L\leq 64$ ($L\geq 160$), used to extract the effective dynamical exponent $z$ for that range of $L$. 
  }
  \label{fig:cross_over}
\end{figure}

To further quantify the above picture, we follow \cite{Crowley2019hyperuniform} by defining a cutoff density $a\rho_0$, with $0<a<1$. We then compute the probability for the density of an $L$-long segment $\rho$ to be smaller than the cutoff, $P_L[\rho<a\rho_0]$. For random disorder, it is expected to decay exponentially $ P_L[\rho<a\rho_0] \sim \exp\qty(-b L)$ for a positive $b$ that depends on the cutoff. In \cref{subfig:rareeventsdecay_a}, we take $a=0.92$, and find that $P_L[\rho<a\rho_0]$ decays in the QP case faster than exponential. In addition, in \cref{subfig:rareeventsdecay_b}, we show that $P_L[\rho<a\rho_0]$ identically vanishes for $L	\gtrsim 180$, indicating that (at least for the number of samples considered) such rare events do not occur. The dependence of the maximal $L$ value above which $P_L[\rho<a\rho_0]$ vanishes on $a$ is presented in \cref{subfig:rareeventsdecay_c}.

From the above reasoning, due to the absence of rare events, it is likely that Griffiths singularities do not play a role in the QP structures considered in this work. Numerically, we have found a moderate increase in the value of the effective dynamical exponent $z$ upon approach to criticality, see \cref{subfig:cross_over_a}. We assume a power-law relation $t\sim L^z$ and fit the entanglement entropy saturation time in the volume-law phase and the ancilla qubit purification time in the area-law phase. In our fit, we consider chain lengths $L\geq 64$. In proximity to the critical point, when the correlation length is comparable to or larger than the chain length, the scaling behavior is controlled by the critical fan. In \cref{subfig:cross_over_b}, we exemplify this cross-over effect by fixing $p=0.212>p_c$ and plotting the ancilla purification time as a function of the chain length $L$. Assuming the same power-law scaling form. We extract the dynamical exponent $z$ for a short ($L\le64$ with $\qty(p-p_c)L^{1/\nu}< 0.6$) versus long ($L\ge160$ with $\qty(p-p_c)L^{1/\nu}> 0.9$) chain. We indeed find that $z$ flows from $0.92$ to $0.58$ as we move away from the critical fan at longer chains. It would be interesting to study the precise scaling across the critical fan in future work.

\bibliography{non_pisot_mipt_prb}

\end{document}